\documentclass[prd,aps,tightenlines,preprintnumbers,amsmath,amssymb,showpacs]
{revtex4}
\usepackage{graphicx}
\usepackage{dcolumn}
\usepackage{bm}
\usepackage{hyperref}

\def \beq{\begin{equation}}
\def \eeq{\end{equation}}

\def\lsim{\mathrel{\rlap{\lower4pt\hbox{\hskip1pt$\sim$}}
    \raise1pt\hbox{$<$}}}                % less than or approx. symbol
\def\gsim{\mathrel{\rlap{\lower4pt\hbox{\hskip1pt$\sim$}}
    \raise1pt\hbox{$>$}}}                % greater than or approx. symbol

\begin{document}

\date{\today}
\preprint{EFI 09-33}

\title{Contributions to the Muon's Anomalous Magnetic Moment from a Hidden 
Sector}
\author{David McKeen}
\email{mckeen@theory.uchicago.edu}
\affiliation{Enrico Fermi Institute and Department of Physics, University of 
Chicago, 5640 South Ellis Avenue, Chicago, IL 60637}
\date{\today}

\begin{abstract}
The measurement of the anomalous magnetic moment of the muon provides a 
stringent test of the standard model and of any physics that lies beyond it.  
There is currently a deviation of $3.1\sigma$ between the standard model 
prediction for the muon's anomalous magnetic moment and its experimental value.  
We calculate the contribution to the anomalous magnetic moment in theories 
where the muon couples to a particle in a hidden sector (that is, uncharged 
under the standard model) and a connector (which has nontrivial standard model gauge and hidden sector quantum numbers).
\end{abstract}

\maketitle

%%%%%%%%%%%%%%%%%%%%%%%%%%%%%%%%%%%%%%%%%%%%%%%%%%%%%%%%%%%%%%%%%%%%%%%%%%%%%%%

\begin{section}{Introduction}
Quantum field theory predicts that the gyromagnetic ratio of the muon will 
differ slightly from its tree-level value of $g_\mu=2$.  Properly accounting 
for the nonzero value of the anomalous magnetic moment, $a_\mu=(g_\mu-2)/2$, of 
the muon is a precise test of the standard model (SM) and of physics beyond the 
SM.

The most recent determination of $a_\mu$ in the SM is \cite{Davier:2009zi}
\begin{align}
a_\mu^{\rm SM}=\left(11~659~183.4\pm 4.9\right)\times10^{-10}~~~.
\label{eq:a_th}
\end{align}
The dominant sources of uncertainty in this expression are the leading-order hadronic vacuum polarization contribution and the contribution from hadronic light-by-light scattering.  In Ref.~\cite{Davier:2009zi}, the leading-order hadronic contribution is determined to be
\begin{align}
a_\mu^{\rm LO~Had.}=\left(695.5\pm 4.1\right)\times10^{-10}~~~,
\end{align}
while the most recent determination of the hadronic light-by-light contribution is \cite{Prades:2009tw}
\begin{align}
a_\mu^{\rm Had.~LbL}=\left(10.5\pm 2.6\right)\times10^{-10}~~~.
\end{align}
The total SM prediction for $a_\mu$ in Eq.~\ref{eq:a_th} differs from the experimental value \cite{Bennett:2006fi},
\begin{align}
a_\mu^{\rm Exp}=\left(11~659~208.0\pm 5.4\pm 3.3\right)\times10^{-10}~~~,
\end{align}
at the $3.1\sigma$ level.  There is some discrepancy in using $e^+e^-$ or 
$\tau$ decay data to extract the leading-order hadronic contribution to $a_\mu$ with $\tau$ 
decay data leading to a $1.9\sigma$ difference between the SM and experimental 
values of $a_\mu$.  For recent reviews of the status of $a_\mu$, see Ref.~\cite{Prades:2009qp}

The difference between $a_\mu^{\rm SM}$ and $a_\mu^{\rm Exp}$ has spurred 
numerous  studies of new physics scenarios that could offer an explanation, for example, supersymmetry \cite{Domingo:2008bb}, universal extra dimensions \cite{Appelquist:2001jz}, and unparticles \cite{Conley:2008jg}.  Another scenario that 
has received attention in the literature is that of a hidden $U(1)^{\prime}$ 
whose gauge boson kinetically mixes with the photon \cite{Holdom:1985ag}.  The 
constraints from $a_\mu$ on such a scenario are discussed in 
\cite{Pospelov:2008zw}.

In this paper we investigate and catalogue the contributions to $a_\mu$ that 
arise from the muon coupling to some hidden sector.  We do this in four 
situations that differ in the spin of the hidden sector particle that couples 
to the muon, and in the spin of other particles present in the interaction to 
preserve gauge invariance.  These scenarios are generalizations of some models 
already investigated, like that of \cite{Pospelov:2008zw}.

Schematically, the interactions we consider are of the form
\begin{align}
{\cal L}_{\rm int}\sim \lambda X Y \mu~~~,
\label{eq:schem_lag}
\end{align}
where Lorentz and gauge indices have been suppressed.  In this Lagrangian and 
in the rest of this work, $X$ refers to a SM singlet that could be charged 
under some hidden symmetry group, which we denote by $G$, and $Y$ is a particle 
that is charged under the SM (to preserve the SM gauge invariance of the 
interaction) and under $G$ if $X$ is (to preserve $G$ invariance).  The 
particles in Eq.~\ref{eq:schem_lag} are classified in the table below: 
\begin{center}
\begin{tabular}{c c c c} \hline \hline
Type of matter & Std.\ Model &    G    & Example \\ \hline
Ordinary       & Non-singlet & Singlet & $\mu$ \\
Connector          & Non-singlet & Non-singlet & $Y$ \\
Hidden         & Singlet     & Non-singlet & $X$ \\ \hline \hline
\end{tabular}
\end{center}
$\lambda$ is the coupling strength of this interaction between the muon, the 
hidden sector particle $X$, and the connector $Y$.   Interactions of this form 
generate corrections to $a_\mu$ of order $\lambda^2$.

We note that $X$ could be a dark matter candidate.  If $m_X<m_Y$ and $X$ is 
the lightest particle with some hidden charge, it could be long lived.  
Indeed, the relic density of $X$ could naturally be driven to the observed 
value of $\Omega_X\simeq 0.23$ although its mass is unconnected to the 
electroweak scale in a WIMPless dark matter scenario \cite{Feng:2008ya}.  For 
$X$ to be a viable dark matter candidate, it cannot be coupled too strongly to 
the SM; that is $\lambda \lsim g_{\rm weak}$.  Of course, this condition is 
relaxed if we do not require that $X$ comprise the most of the dark matter 
density.  These scenarios have been studied in situations where $X$ couples to 
$b$ quarks, leading to an explanation of the DAMA/LIBRA signal 
\cite{Feng:2008dz} and to missing energy in decays of mesons with $b$ quarks 
\cite{McKeen:2009rm}.

In Sec.~\ref{sec:collconstr}, we discuss constraints on $X$ and $Y$ from collider experiments.  
In Sec.~\ref{sec:calc}, we present the contributions to $a_\mu$ due to several 
scenarios of the form of Eq.~\ref{eq:schem_lag}.  We discuss 
constraints from the measured value of $a_\mu$ on these scenarios in Sec.~\ref{sec:exp}, and, in Sec.~\ref{sec:conclusion}, we conclude.
\end{section}

%%%%%%%%%%%%%%%%%%%%%%%%%%%%%%%%%%%%%%%%%%%%%%%%%%%%%%%%%%%%%%%%%%%%%%%%%%%%%%%

\begin{section}{Collider constraints on $X$ and $Y$}\label{sec:collconstr}
If $X$ is a SM singlet that is only weakly coupled to the SM, as we assume 
here, then there are no firm constraints on its allowed mass coming from 
collider experiments.  We consider its mass to be essentially free in this study.

There are, however, tight bounds on the possible mass of $Y$ since it has the 
same electric charge as the muon.  The firmest bounds come from the LEP 
experiments' searches for right-handed sleptons.  These experiments looked for a pair of 
sleptons produced by a virtual photon or $Z$ that decay to a pair of acoplanar 
leptons along with two neutralinos (missing energy).  Such searches apply in 
the case of a Lagrangian of the form of Eq.~\ref{eq:schem_lag} if 
$m_X<m_Y-m_\mu$ and $\lambda$ large enough that the $Y$'s decay promptly,
that is, $\lambda\gsim 10^{-8}$.

The ALEPH, DELPHI, L3, and OPAL experiments set a combined limit 
\cite{Heister:2003zk} on the production of smuons decaying to muons and missing 
energy of
\begin{align}
\sigma\left(e^+e^-\to \tilde{\mu}_R \bar{\tilde{\mu}}_R\right)<0.08~{\rm pb}~~~
{\rm at}~95\%~{\rm C.L.}
\end{align}
if ${\cal B}\left(\tilde{\mu}_R\to \chi^0_1 \mu^-\right)=1$ for 
$m_{\tilde{\mu}_R}\lsim 95~{\rm GeV}$ and $m_{\tilde{\mu}_R}-m_{\chi^0_1}
\gsim 10~{\rm GeV}$ at a rescaled center-of-mass energy $\sqrt{s}=208~{\rm GeV}$.  If 
$m_X\lsim m_Y +10~{\rm GeV}$ and $Y\to X\mu$ is the dominant decay mode for 
$Y$, then this limit should also hold for $Y$ pair production, 
assuming acceptances don't differ too drastically.

In this paper we will consider a fermionic $Y$ in Secs.~\ref{sec:case1} and~\ref{sec:case2}, a scalar $Y$ in Sec. \ref{sec:case3}, and a vector $Y$ in \ref{sec:case4}.  In 
any of these cases, $Y$ has the same electric charge as the muon since $X$ is assumed to be electrically
neutral.  Its charges under electroweak $SU(2)_L\times U(1)_Y$ depend on whether 
the interaction of Eq.~\ref{eq:schem_lag} respects electroweak symmetry.  Of course, 
what electroweak charges we assign to $Y$ are important in estimating 
the production cross section at LEP.

In the case of a fermionic $Y$, the simplest case is that of a heavy chrial lepton whose $SU(2)_L\times U(1)_Y$ charges are the same as that of the muon.  The production cross section, $\sigma\left(e^+e^-\to\gamma^* Z^*\to Y\bar{Y}\right)$, is calculated in Sec.~\ref{sec:fermprod} and is plotted in Fig.~\ref{fig:cross} (a).

A scalar $Y$ could either couple to $\mu_L$ or $\mu_R$.  We label each of these as $Y_L$ and $Y_R$ respectively, where the subscript does not indicate any chirality for $Y$ since it has none, but the chirality of the muon to which it couples.  This is the situation with sleptons where, for example, $\tilde{\mu}_L$ and $\tilde{\mu}_R$ are different states.  $Y_L$ and $Y_R$ each have unit electric charge which fixes their couplings to photons.  We also choose that $Y_L$ couples in a gauge invariant way to $Z$ bosons with the same strength as $\mu_L$ and similarly for $Y_R$.  The production cross sections, $\sigma\left(e^+e^-\to\gamma^* Z^*\to Y_L^+Y_L^-\right)$ and $\sigma\left(e^+e^-\to\gamma^* Z^*\to Y_R^+Y_R^-\right)$, are derived in Sec.~\ref{sec:scalarprod} and are plotted in Fig.~\ref{fig:cross} (b).

The situation where $Y$ is a vector boson is more complicated as further states need to be introduced to maintain unitarity.  As in the scalar case, there are again two $Y$s which we label in terms of the handedness of the muon that they couple to, $Y_L^\nu$ and $Y_R^\nu$.  These vector bosons are electrically charged which again fixes their coupling to photons.  If this is the only coupling that contributes to $Y$ pair production, then the cross section $\sigma\left(e^+e^-\to\gamma^*\to Y_{L, R}^{\nu+}Y_{L, R}^{\nu-}\right)$ diverges as the center-of-mass energy increases, in conflict with unitarity.  Only adding in a coupling of $Y_{L, R}^{\nu}$ to the $Z$ does not fix this since the $Z$ has a chiral coupling to leptons while the photon's is vector-like.  This is the same problem faced when calculating $\sigma\left(e^+e^-\to W^{\nu+}W^{\nu-}\right)$.  The solution there is to include $t$-channel neutrino exchange in addition to $s$-channel photon and $Z$ exchange.  We consider the case where the solution to the unitarity problem in vector $Y$ pair production is similar; we assume that there are fermions, $N_L$ and $N_R$, which are electrically neutral that are exchanged in the $t$-channel.  This is the case if, for example, $Y_R^\nu$ is a heavy charged gauge boson associated with a broken $SU(2)_R$ and $N_R$ is a right-handed neutrino.  A similar situation occurs in little Higgs models with T-Parity where we can consider $Y_L^\nu$ as a T-odd vector boson and $N_L$ as a T-odd neutrino.  $N_L$ or $N_R$ could also be thought of as the singlet in an interaction of the form of that in Eq.~\ref{eq:schem_lag} with the muon replaced by the electron.  In any one of these scenarios, the requirement that the production cross section eventually vanishes as the center-of-mass energy grows implies some relationships between the couplings of $Y_{L,R}^\nu$ to the $Z$ and to $e_{L,R}-N_{L,R}$.  The cross sections $\sigma\left(e^+e^-\to Y_{L}^{\nu+}Y_{L}^{\nu-}\right)$ and $\sigma\left(e^+e^-\to Y_{R}^{\nu+}Y_{R}^{\nu-}\right)$ are calculated in Sec.~\ref{sec:vectorprod} and are shown for different masses of $N_L$ and $N_R$ in Figs.~\ref{fig:cross} (c) and (d).  Since we do not assume anything about the coupling of $Y_{L,R}^{\nu}$ to quarks, the stringent limits on heavy charged vector bosons from hadron colliders are ignored.

In Fig.~\ref{fig:cross}, it is seen that the $Y^+Y^-$ production cross section is 
greater than $0.08~{\rm pb}$ for $m_Y\gsim 89~{\rm GeV}$ in each of these 
cases.  If $Y$ is long-lived on detector time scales ($m_X$ could be larger 
than $m_Y$ or $\lambda \lsim 10^{-8}$) then tracks would have been seen in the 
electromagnetic calorimeters in the LEP experiments as 
long as the center-of-mass energy was above $Y$ threshold.  We consider this scenario to 
be ruled out.
\begin{figure*}
\begin{center}
\begin{tabular}{cc}
\resizebox{80mm}{!}{\includegraphics{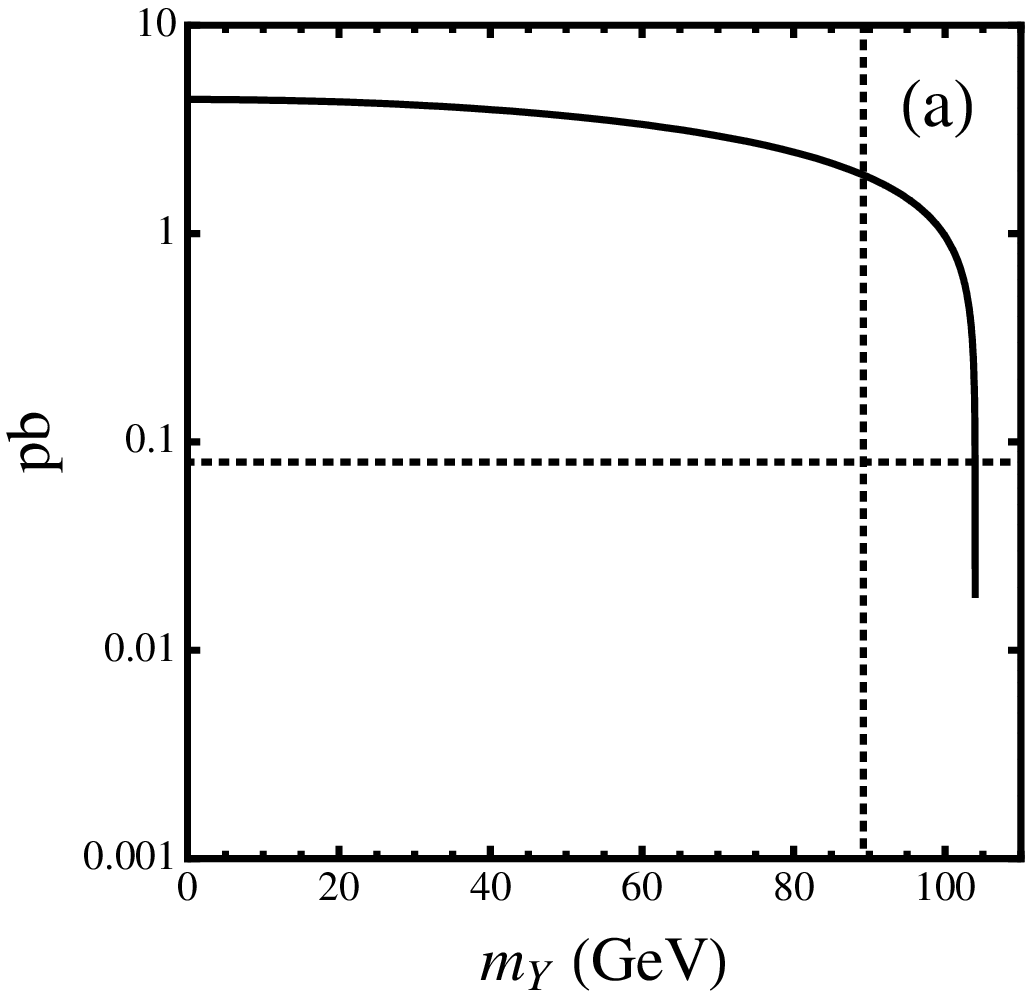}} &
\resizebox{80mm}{!}{\includegraphics{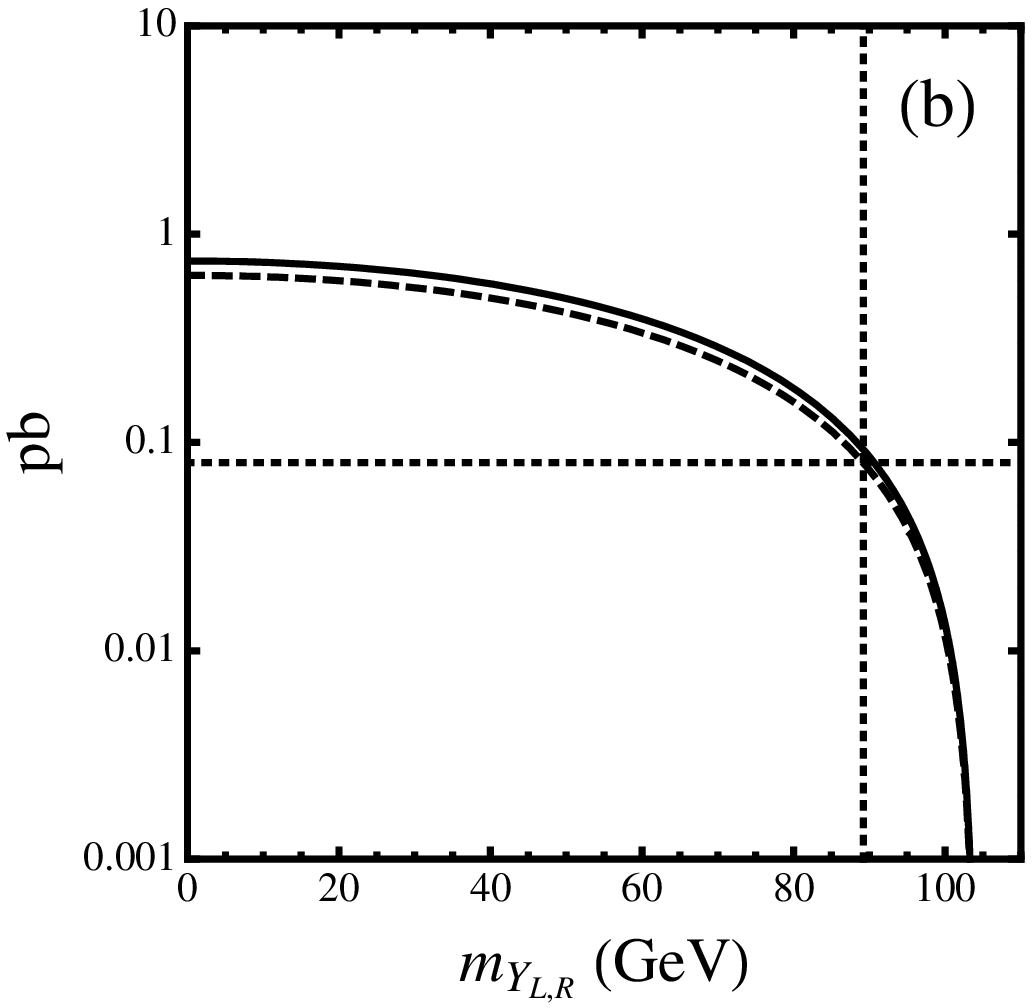}} \\
\resizebox{80mm}{!}{\includegraphics{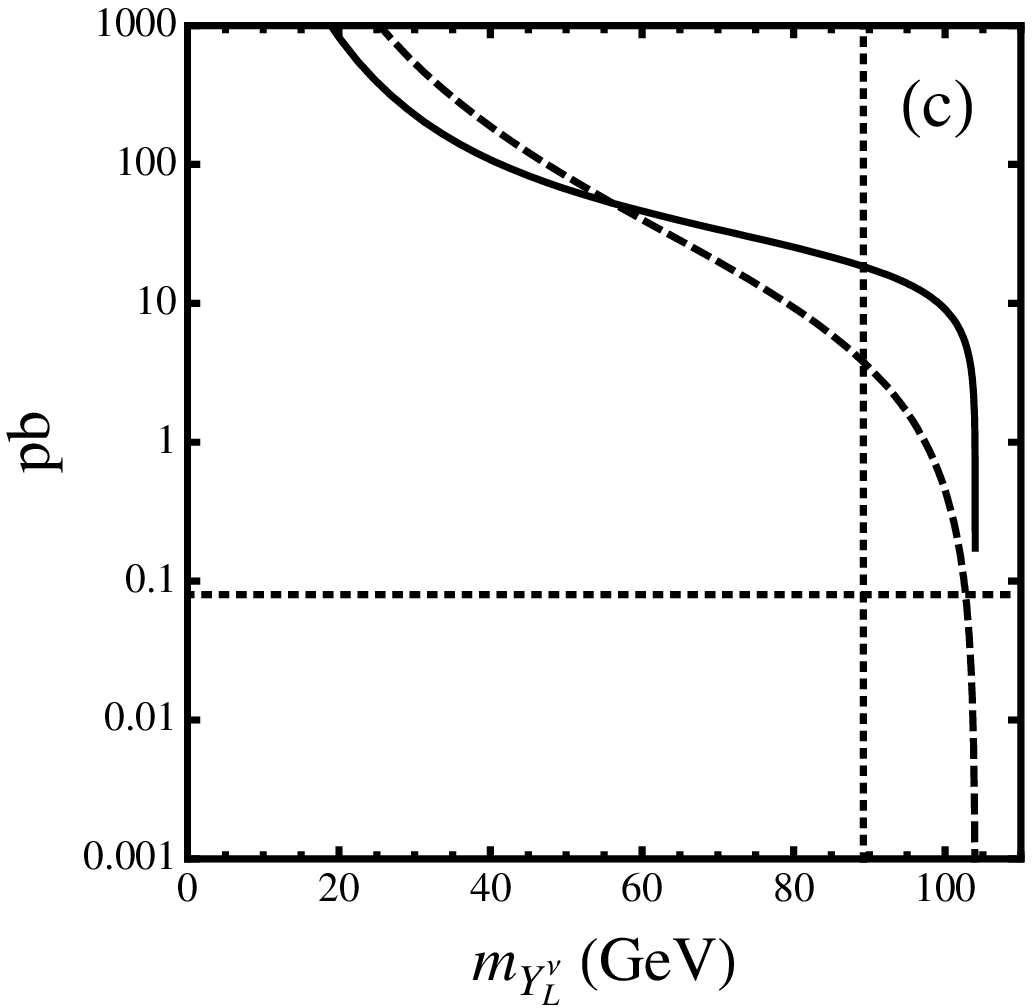}} &
\resizebox{80mm}{!}{\includegraphics{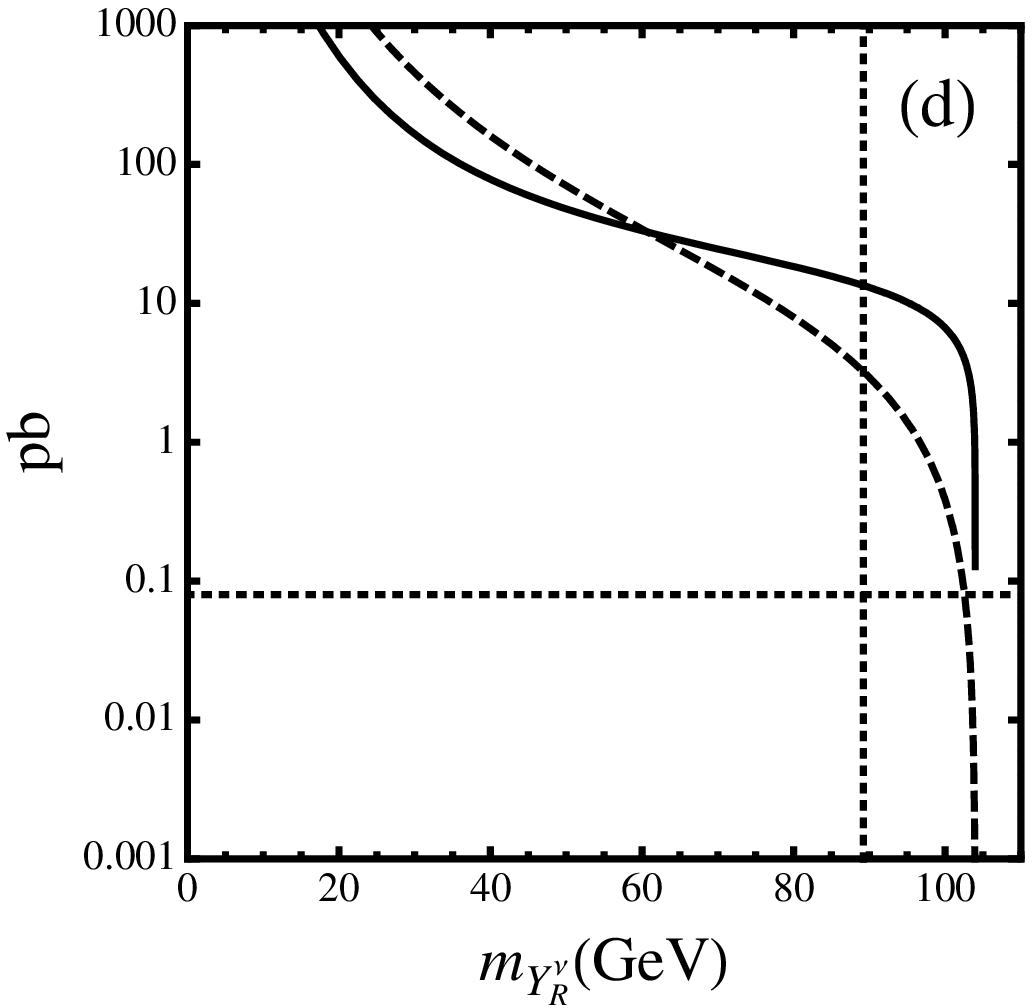}}
\end{tabular}
\caption{(a) $\sigma\left(e^+e^-\to Y\bar{Y}\right)$ at $\sqrt{s}=208~{\rm GeV}$ where $Y$ is a fermion.  (b) $\sigma\left(e^+e^-\to Y_{L}^+Y_{L}^-\right)$ (solid) and $\sigma\left(e^+e^-\to Y_{R}^+Y_{R}^-\right)$ (dashed) at $\sqrt{s}=208~{\rm GeV}$ where $Y$ is a scalar.  (c) $\sigma\left(e^+e^-\to Y_{L}^{\nu+}Y_{L}^{\nu-}\right)$ where $Y$ is a vector boson for $m_{N_L}=0$ (solid) and $m_{N_L}=10~{\rm TeV}$ (dashed).  (c) $\sigma\left(e^+e^-\to Y_{R}^{\nu+}Y_{R}^{\nu-}\right)$ where $Y$ is a vector boson for $m_{N_R}=0$ (solid) and $m_{N_R}=10~{\rm TeV}$ (dashed).  The horizontal dotted lines in each plot indicate the LEP limit of $0.08~{\rm pb}$ and the vertical dotted lines indicate the lower bound on $m_Y$ of $89~{\rm GeV}$ which comes from scalar $Y_R$ pair production as seen in (b).  Note the $p$-wave suppression of the production of a scalar $Y$ pair in (b) near threshold which causes its cross section to decrease more steadily as a function of increasing $m_Y$ than the cuspier cross section for fermionic $Y$ pair production in (a).  We also see that for $m_{N_{L,R}}=0$ in (c) and (d), there is no $p$-wave suppression of the cross section of a vector $Y$ pair near threshold whereas when we decouple $N_{L,R}$ by taking its mass to $10~{\rm TeV}$, there is a $p$-wave suppression.  This suppression can be seen in the expression for the production cross section in Sec.~\ref{sec:vectorprod}; for  $m_{N_{L,R}}\gg \sqrt{s}$, the cross section is proportional to $\beta^3$.}
\label{fig:cross}
\end{center}
\end{figure*}

The situation is complicated if there are neutrinos with masses above $m_Z/2$ 
that are part of an $SU(2)_L$ doublet with $Y_L$ or if lepton family violating 
decays compete with $Y\to X\mu$.  However, searches for acoplanar $e^+e^-$ or 
$\tau^+\tau^-$ pairs and missing energy yield similar limits on the production 
cross section of selectrons and staus.  Therefore, in this work, we take a lower bound of 
$m_Y \gsim 89~{\rm GeV}$.

If $Y$ only receives SM contributions to its mass perturbativity could become an issue if 
$m_Y \gsim 500~{\rm GeV}$.  We do not explore this issue in detail.
\end{section}

%%%%%%%%%%%%%%%%%%%%%%%%%%%%%%%%%%%%%%%%%%%%%%%%%%%%%%%%%%%%%%%%%%%%%%%%%%%%%%%

\begin{section}{Contributions to $a_\mu$ due to interaction of the 
muon with a hidden sector}\label{sec:calc}
In this paper we investigate the consequences of the muon coupling to a standard 
model singlet, which we denote by $X$, and to a particle charged under the 
standard model, which we call $Y$.  There are four cases we consider based on 
the intrinsic angular momenta of $X$ and $Y$.  The first case is a spin-0 $X$ 
and a spin-1/2 $Y$.  The second is a spin-1 $X$ and a spin-1/2 $Y$.  The third 
case is a spin-1/2 $X$ and a spin-0 $Y$ while the last is a spin-1/2 $X$ and a 
spin-1 $Y$.  We present the contributions to $a_\mu$ in each case below.

%%%%%%%%%%%%%%%%%%%%%%%%%%%%%%%%%%%%%%%%%%%%%%%%%%%%%%%%%%%%%%%%%%%%%%%%%%%%%%%

\begin{subsection}{Case I}\label{sec:case1}
In the first case, the interaction Lagrangian is given by
\begin{align}
%{\cal L}_{\rm int}=\lambda X \bar{Y} \mu + {\rm H.c.}~~~.
{\cal L}_{\rm int}=\lambda_L X \bar{Y}_R \mu_L
+\lambda_R X \bar{Y}_L \mu_R + {\rm H.c.}~~~.
\end{align}
This contributes to the muon's anomalous magnetic moment through the 
diagram seen in Fig.~\ref{fig:diagrams} (a).  This contribution is easily calculated to 
be
\begin{align}
\left(\Delta a_\mu\right)_{1}&=\frac{1}{16\pi^2}\int_0^1 dx \frac{\left(1-x\right)^2
\left[\left(\lambda_L^2+\lambda_R^2\right)m_\mu m_Y+
2\lambda_L \lambda_R xm_\mu^2\right]}{(1-x)m_Y^2+xm_X^2-x(1-x)m_\mu^2}~~~.
\end{align}
If $m_Y,m_X\gg m_\mu$ then we can approximate this expression as
\begin{align}
\left(\Delta a_\mu\right)_{1}&\simeq\frac{1}{16\pi^2}\left(\lambda_L^2+\lambda_R^2\right) 
\int_0^1 dx \frac{\left(1-x\right)^2m_\mu m_Y}{(1-x)m_Y^2+xm_X^2}
\\
&=\frac{1}{32\pi^2}\left(\lambda_L^2+\lambda_R^2\right)\frac{m_\mu}{m_Y}
H_1\left(\frac{m_X^2}{m_Y^2}\right)
\\
&=8.36\times 10^{-7}\left(\lambda_L^2+\lambda_R^2\right) 
\left(\frac{400~\rm GeV}{m_Y}\right)H_1\left(\frac{m_X^2}{m_Y^2}\right)~~~,
\end{align}
where
\begin{align}
H_1\left(r\right)&=2\int_0^1 dx 
\frac{\left(1-x\right)^2}{1-\left(1-r\right)x}~~~.
\end{align}
\begin{figure*}
\begin{center}
\begin{tabular}{ccc}
\includegraphics{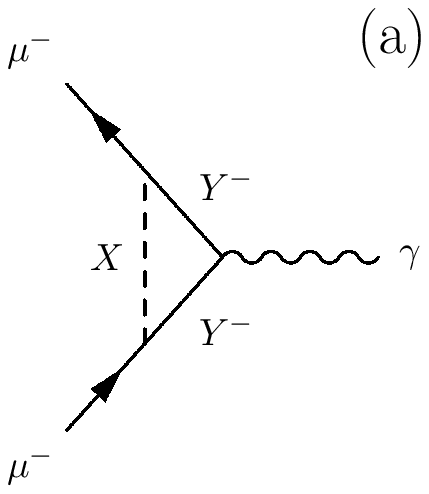} & ~~~~~~ &
\includegraphics{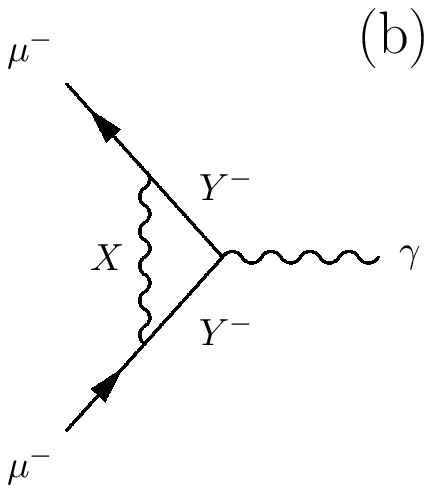} \\ \\
\includegraphics{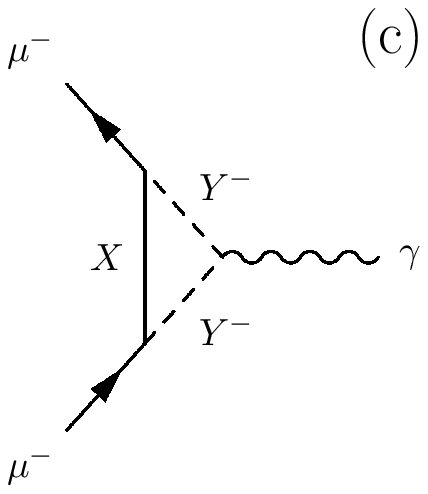}  & ~~~~~~ &
\includegraphics{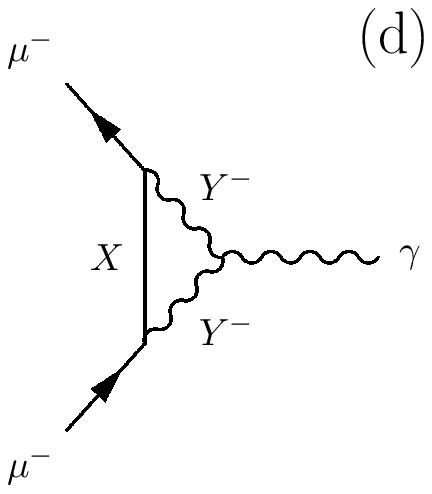}
\end{tabular}
\caption{Diagrams relevant for Cases I (a), II (b), III (c), IV (d).}
\label{fig:diagrams}
\end{center}
\end{figure*}
\end{subsection}

%%%%%%%%%%%%%%%%%%%%%%%%%%%%%%%%%%%%%%%%%%%%%%%%%%%%%%%%%%%%%%%%%%%%%%%%%%%%%%%

\begin{subsection}{Case II}\label{sec:case2}
In the second case, the interaction Lagrangian is now given by
\begin{align}
%{\cal L}_{\rm int}=\lambda X^{\mu}\bar{Y}\gamma_{\mu} \mu + {\rm H.c.}~~~.
{\cal L}_{\rm int}=\lambda_L X^{\mu}\bar{Y}_L\gamma_{\mu} \mu_L
+\lambda_R X^{\mu}\bar{Y}_R\gamma_{\mu} \mu_R + {\rm H.c.}~~~.
\end{align}
This gives a contribution to $a_\mu$ through the diagram seen in Fig.~
\ref{fig:diagrams} (b).  We find
%\begin{align}
%\left(\Delta a_\mu\right)_{2}&=\frac{1}{8\pi^2}\int_0^1 dx 
%\frac{x\left(1-x\right)\left(4\lambda_L \lambda_R m_\mu m_Y-
%\left(\lambda_L^2+\lambda_R^2\right) \left(1+x\right)m_\mu^2\right)}
%{(1-x)m_Y^2+xm_X^2-x(1-x)m_\mu^2}
%\end{align}
\begin{align}
\left(\Delta a_\mu\right)_{2}&=\frac{1}{8\pi^2}\int_0^1 dx 
\frac{x\left(1-x\right)\left[4\lambda_L \lambda_R m_\mu m_Y-
\left(\lambda_L^2+\lambda_R^2\right) \left(1+x\right)m_\mu^2\right]}
{(1-x)m_Y^2+xm_X^2-x(1-x)m_\mu^2}
\\
&+\frac{1}{16\pi^2}\frac{m_\mu^2}{m_X^2}\int_0^1 dx 
\frac{\left(1-x\right)^3\left[
2\lambda_L \lambda_R\left(1-x\right) m_\mu m_Y-\left(\lambda_L^2+\lambda_R^2\right) \left(m_Y^2-xm_\mu^2\right)\right]}
{(1-x)m_Y^2+xm_X^2-x(1-x)m_\mu^2}
\end{align}
%\begin{align}
%\left(\Delta a_\mu\right)_{2}&=\frac{1}{16\pi^2}\int_0^1 dx 
%\frac{x\left(1-x\right)\left[2\lambda_L \lambda_R\left(4+\left(1-x\right)^3\left(m_\mu^2/m_X^2\right)\right) m_\mu m_Y-
%\left(\lambda_L^2+\lambda_R^2\right) \left(1+x\right)m_\mu^2\right]}
%{(1-x)m_Y^2+xm_X^2-x(1-x)m_\mu^2}
%\\
%&-\frac{1}{16\pi^2}\frac{m_\mu^2}{m_X^2}\int_0^1 dx 
%\frac{\left(1-x\right)^3\left(
%\left(\lambda_L^2+\lambda_R^2\right) \left(m_Y^2-xm_\mu^2\right)-2\lambda_L \lambda_R\left(1-x\right) m_\mu m_Y\right)}
%{(1-x)m_Y^2+xm_X^2-x(1-x)m_\mu^2}
%\end{align}
If $m_Y,m_X\gg m_\mu$,
\begin{align}
\left(\Delta a_\mu\right)_{2}&\simeq\frac{m_\mu m_Y}{2\pi^2}\lambda_L \lambda_R \int_0^1 dx 
\frac{x\left(1-x\right)}{(1-x)m_Y^2+xm_X^2}\nonumber
\\
&~~~~-\frac{m_\mu^2 m_Y^2}{16\pi^2 m_X^2}\left(\lambda_L^2+\lambda_R^2\right) \int_0^1 dx 
\frac{\left(1-x\right)^3}{(1-x)m_Y^2+xm_X^2}
\\
&=\frac{1}{4\pi^2}\lambda_L \lambda_R\left(\frac{m_\mu}{m_Y}\right)
H_2\left(\frac{m_X^2}{m_Y^2}\right)
\\
&~~~~-\frac{1}{48\pi^2}\left(\lambda_L^2+\lambda_R^2\right) \frac{m_\mu^2}{m_X^2}G_2\left(\frac{m_X^2}{m_Y^2}\right)
\\
&=6.69\times 10^{-6}\lambda_L \lambda_R\left(\frac{400~\rm GeV}{m_Y}\right)
H_2\left(\frac{m_X^2}{m_Y^2}\right)
\\
&~~~~-2.36\times 10^{-5}\left(\lambda_L^2+\lambda_R^2\right)\left(\frac{1~\rm GeV}{m_X}\right)^2
H_2\left(\frac{m_X^2}{m_Y^2}\right)~~~,
\end{align}
where
\begin{align}
H_2\left(r\right)&=2\int_0^1 dx 
\frac{x\left(1-x\right)}{1-\left(1-r\right)x}~~~,
\\
G_2\left(r\right)&=3\int_0^1 dx 
\frac{\left(1-x\right)^3}{1-\left(1-r\right)x}~~~,
\label{eq:h2}
\end{align}

This interaction is a generalization of the much-discussed case in which the photon 
kinetically mixes with a $\rm GeV$ scale gauge boson.  To obtain the 
contribution to the muon's anomalous magnetic moment in this situation, we 
identify $Y$ with the muon and write 
$\lambda_L=\lambda_R=\epsilon e$ where $\epsilon$ characterizes the strength 
of the kinetic mixing and $e$ is the strength of the muon's electric charge.  
Then (as in \cite{Pospelov:2008zw}),
\begin{align}
\left(\Delta a_\mu\right)_{2\prime}&=\frac{\epsilon^2\alpha  m_\mu^2}{\pi}
\int_0^1 dx \frac{x\left(1-x\right)^2}
{(1-x)^2 m_\mu^2+xm_X^2}
\end{align}
If $m_X\gg m_\mu$ we can approximate this as
\begin{align}
\left(\Delta a_\mu\right)_{2 \prime}&\simeq\frac{\epsilon^2 \alpha}{3\pi}
\left(\frac{m_\mu}{m_X}\right)^{2}
\\
&=8.65\times 10^{-6}\epsilon^2\left(\frac{1~\rm GeV}{m_X}\right)^{2}~~~,
\end{align}
while if $m_X\ll m_\mu$,
\begin{align}
\left(\Delta a_\mu\right)_{2 \prime}&\simeq\frac{\epsilon^2 \alpha}{2\pi}
\\
&=1.16\times 10^{-3}\epsilon^2~~~.
\end{align}
These expressions agree with those in Ref.~\cite{Pospelov:2008zw}.
\end{subsection}

%%%%%%%%%%%%%%%%%%%%%%%%%%%%%%%%%%%%%%%%%%%%%%%%%%%%%%%%%%%%%%%%%%%%%%%%%%%%%%%

\begin{subsection}{Case III}\label{sec:case3}
$X$ is now a fermion, while $Y$ is a scalar.  The interaction is given by
\begin{align}
%{\cal L}_{\rm int}=\lambda Y \bar{X} \mu + {\rm H.c.}~~~.
{\cal L}_{\rm int}=\lambda_L Y_L \bar{X} \mu_L 
+\lambda_R Y_R \bar{X} \mu_R + {\rm H.c.}~~~.
\end{align}
Here, the subscript on $Y$ labels the helicity of the muon to which it couples and nothing about its own helicity, just as the subscripts that label sfermions in supersymmetry do.  In Cases I and II, $Y_L$ and $Y_R$ were two-component Weyl spinors married to form a Dirac fermion whose mass term breaks electroweak symmetry.  Here, they are separate fields that, in general, have different masses.  The diagram shown in Fig.~\ref{fig:diagrams} (c) gives a contribution 
to $a_\mu$ of
\begin{align}
\left(\Delta a_\mu\right)_{3}&=\frac{\lambda_L^2}{16\pi^2}\int_0^1 dx \frac{x\left(1-x\right)m_\mu m_X}
{(1-x)m_{Y_L}^2+xm_X^2-x(1-x)m_\mu^2}+\left(L\to R\right)~~~.
\end{align}
If $m_Y,m_X\gg m_\mu$ then we can approximate this expression as
\begin{align}
\left(\Delta a_\mu\right)_{3}&\simeq\frac{m_\mu m_X}{32\pi^2}\left[\frac{\lambda_L^2}{m_{Y_L}^2}H_2\left(\frac{m_X^2}{m_{Y_L}^2}\right)+\frac{\lambda_R^2}{m_{Y_R}^2}H_2\left(\frac{m_X^2}{m_{Y_R}^2}\right)\right]
\\
&=2.09\times 10^{-9}\left[\lambda_L^2\left(\frac{400~\rm GeV}{m_{Y_L}}\right)^2H_2\left(\frac{m_X^2}{m_{Y_L}^2}\right)+\lambda_R^2\left(\frac{400~\rm GeV}{m_{Y_R}}\right)^2H_2\left(\frac{m_X^2}{m_{Y_R}^2}\right)\right]
\left(\frac{m_X}{1~\rm GeV}\right)~~~,
\end{align}
where $H_2$ is defined in Eq.~\ref{eq:h2}.
\end{subsection}

%%%%%%%%%%%%%%%%%%%%%%%%%%%%%%%%%%%%%%%%%%%%%%%%%%%%%%%%%%%%%%%%%%%%%%%%%%%%%%%

\begin{subsection}{Case IV}\label{sec:case4}
The last case we consider is a fermionic $X$ and a spin-1 $Y$.  The interaction 
is now
\begin{align}
%{\cal L}_{\rm int}=\lambda Y^\mu \bar{X}\gamma_\mu \mu + {\rm H.c.}~~~.
{\cal L}_{\rm int}=\lambda_L Y_L^\nu \bar{X}\gamma_\nu \mu_L
+\lambda_R Y_R^\nu \bar{X}\gamma_\nu \mu_R + {\rm H.c.}~~~.
\end{align}
As in Case III, the subscript on $Y$ only labels the muon to which it couples.  The relevant diagram is shown in Fig.~\ref{fig:diagrams} (d).  In this case the 
contribution to $a_\mu$ is
\begin{align}
\left(\Delta a_\mu\right)_{4}&=\frac{\lambda_L^2}{8\pi^2}\int_0^1 dx \frac{\left(1-x\right)^2\left(2-x\right)m_\mu^2}{(1-x)m_{Y_L}^2+xm_X^2-x(1-x)m_\mu^2}+{\cal O}\left(\frac{m_\mu^2}{m_{Y_L}^2}\right)+\left(L\to R\right)~~~.
\end{align}
If $m_Y,m_X\gg m_\mu$ then we can approximate this expression as
\begin{align}
\left(\Delta a_\mu\right)_{4}&\simeq\frac{\lambda_L^2}{8\pi^2}\frac{m_\mu^2}{m_{Y_L}^2} \int_0^1 dx 
\frac{\left(1-x\right)^2\left(2-x\right)}{1-x+x\left(m_X^2/m_{Y_L}^2\right)}+\left(L\to R\right)
\\
&=7.36\times 10^{-10}\left[\lambda_L^2\left(\frac{400~\rm GeV}{m_{Y_L}}\right)^2H_4\left(\frac{m_X^2}{m_{Y_L}^2}\right)+\lambda_R^2\left(\frac{400~\rm GeV}{m_{Y_R}}\right)^2H_4\left(\frac{m_X^2}{m_{Y_L}^2}\right)\right]~~~,
\end{align}
where
\begin{align}
H_4\left(r\right)&=\frac{6}{5}\int_0^1 dx 
\frac{\left(1-x\right)^2\left(2-x\right)}{1-\left(1-r\right)x}~~~.
\end{align}
\end{subsection}
\end{section}

%%%%%%%%%%%%%%%%%%%%%%%%%%%%%%%%%%%%%%%%%%%%%%%%%%%%%%%%%%%%%%%%%%%%%%%%%%%%%%%

\begin{section}{Comparison With Experiment}\label{sec:exp}
The deviation of the standard model and experimental values for $a_\mu$ is
\begin{align}
\Delta a_\mu=a_\mu^{\rm Exp}-a_\mu^{\rm SM}=\left(24.6 \pm 8.0\right) \times 
10^{-10}~~~.
\label{eq:exp_mius_th}
\end{align}
%For comparison we note that
%\begin{align}
%a_\mu^{\rm EW}=154 \pm 2 \times 10^{-11}~~~.
%\end{align}
This discrepancy could be lessened if additional sources contribute to the 
muon's anomalous magnetic moment, as in the cases above. In Fig.~\ref{fig:a_mu} 
we plot the contribution to $a_\mu$ in each of the four cases as functions of 
$m_X$ while fixing $\lambda_L=0.1$, $ \lambda _R=0$, and $m_Y=400~{\rm GeV}$ in Cases I and II, and $m_{Y_L}=400~{\rm GeV}$ in Cases III and IV.  In Case II, we have actually plotted $-\left(\Delta a_\mu\right)_2$, since, for these parameter choices, it is negative.
\begin{figure}
\begin{center}
\resizebox{80mm}{!}{\includegraphics{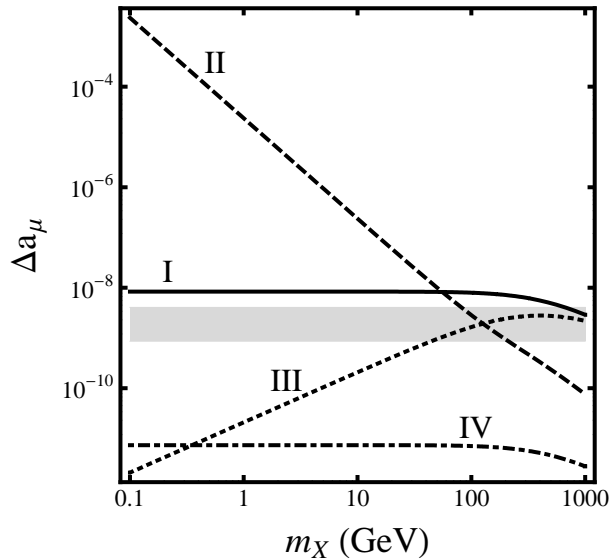}}
\caption{Contributions to $a_\mu$ as functions of $m_X$ for 
$\lambda_L=0.1$, $ \lambda _R=0$, and $m_Y=400~{\rm GeV}$ in Cases I (solid), II 
(dashed), III (dotted) and IV (dot-dashed).  Note that we have plotted $-\left(\Delta a_\mu\right)_2$ in Case II (dashed) since it is negative for these choices of $\lambda_{L,R}$.  We use the full one loop expressions for $\left(\Delta a_\mu\right)_1, \dots, \left(\Delta a_\mu\right)_4$.  The light gray band shows values of 
$\Delta a_\mu$ for which the discrepancy between the theoretical and experimental values of $a_\mu$ (Eq.~\ref{eq:exp_mius_th}) is reduced to $1\sigma$.}
\label{fig:a_mu}
\end{center}
\end{figure}
We see that the helicity flip along the fermion line gives a factor of $m_X$ in 
Case III, which suppresses its contributions to $a_\mu$ at small 
$m_X$ for fixed $m_{Y_L}$.  For smaller values of $m_X$, Case II gives a larger contribution to $a_\mu$ than in 
any of the other scenarios.  We note that the contributions to $a_\mu$ for a fermionic $X$ are generally smaller than for a bosonic $X$, given the same value of the coupling.

If any one of these scenarios describes the dominant contribution to the muon's 
anomalous magnetic moment beyond the standard model, we can ask what values of 
$\lambda_{L,R}$ for a given $m_X$ and $m_Y$ reduce the difference between the 
experimental and theoretical values of $a_\mu$ to less than $2\sigma$.  Fixing $\lambda_R=0$ and
$m_Y=400~{\rm GeV}$ in Cases I and II, and $m_{Y_L}=400~{\rm GeV}$ in Cases III and IV, we plot such values of 
$\lambda_L$ as functions of $m_X$ in 
Fig.~\ref{fig:lambda_allowed}.
\begin{figure*}
\begin{center}
\begin{tabular}{cc}
\resizebox{80mm}{!}{\includegraphics{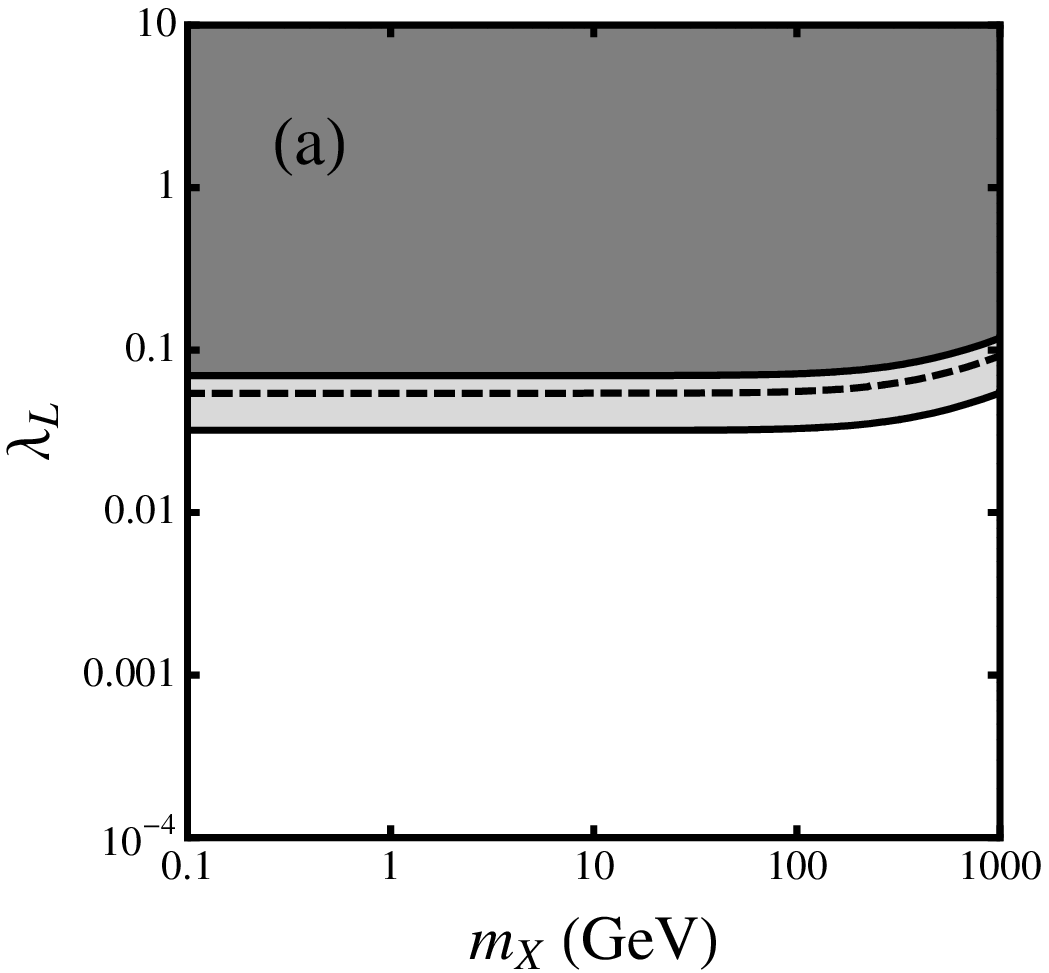}} &
\resizebox{80mm}{!}{\includegraphics{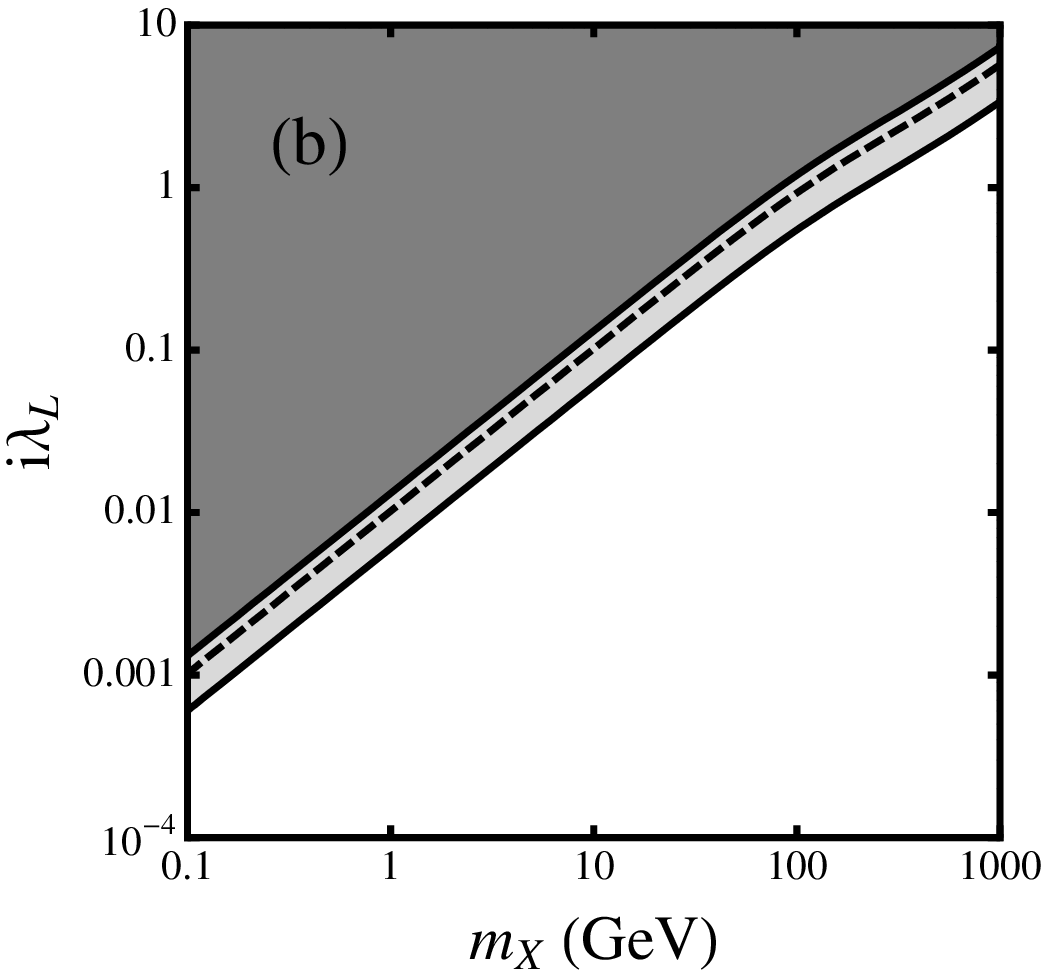}} \\
\resizebox{80mm}{!}{\includegraphics{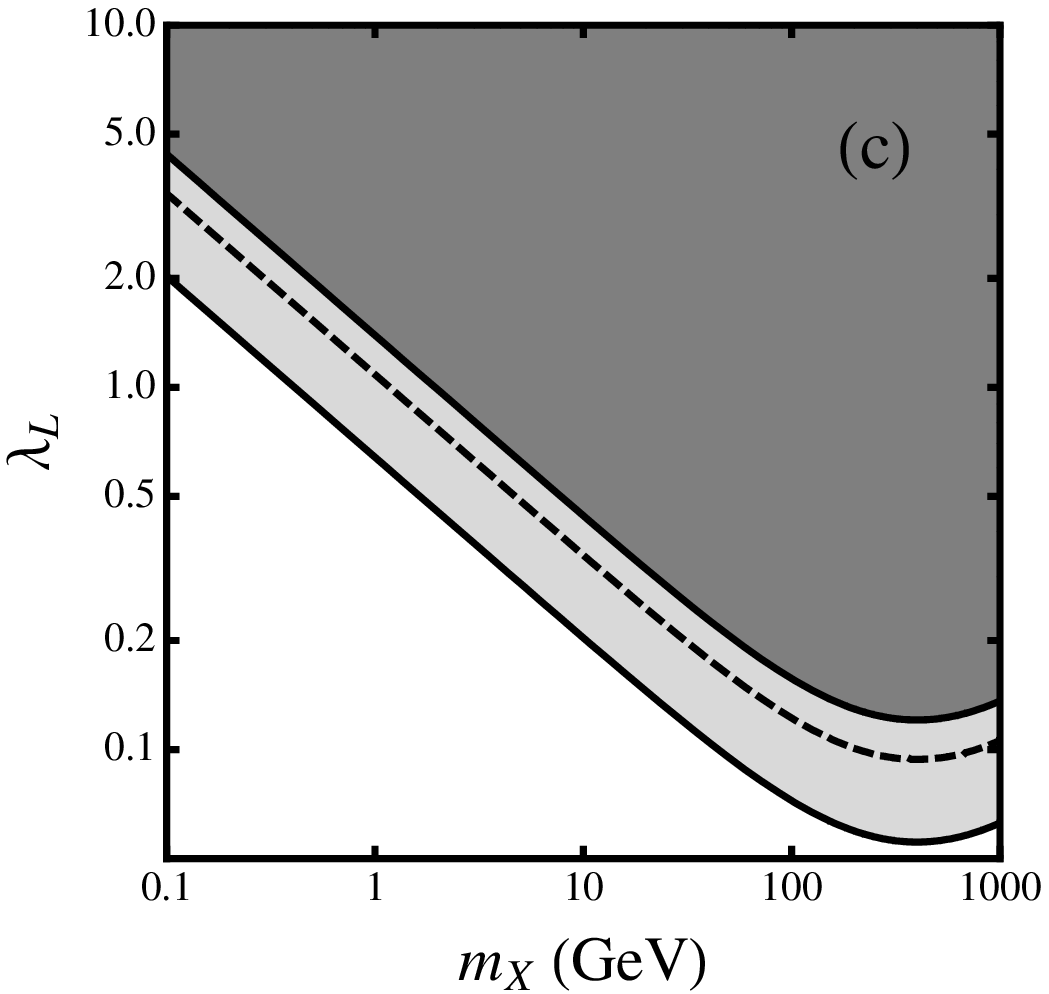}} &
\resizebox{80mm}{!}{\includegraphics{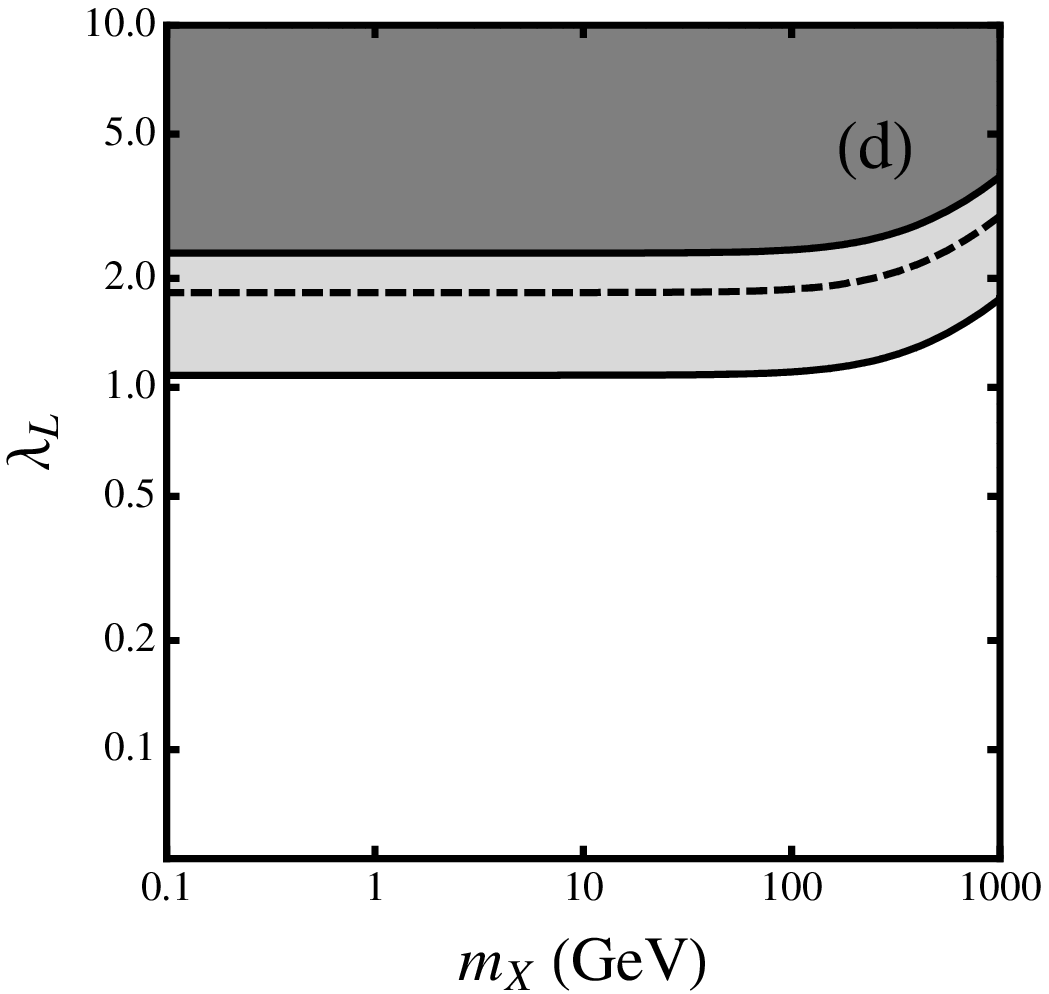}}
\end{tabular}
\caption{Allowed values of $\lambda_L$ in the situation where  
$X$ and $Y$ couple only to $\mu_L$, $\lambda_R=0$, as a 
function of $m_X$ for $m_Y=400~{\rm GeV}$ in Cases I (a) and II (b) and $m_{Y_L}=400~{\rm GeV}$ in Cases III (c) and IV (d).  In Case II (b), we plot $i\lambda_L$ since a real $\lambda_L$  would give a negative $\left(\Delta a_\mu\right)_{2}$.  We use the full one loop expressions for $\left(\Delta a_\mu\right)_1, \dots, \left(\Delta a_\mu\right)_4$.  The light gray bands indicate values of $\lambda$ and $m_X$ for which 
the discrepancy between the theoretical and experimental values of $a_\mu$ is reduced to
less than $2\sigma$.  The dashed lines show the values of $\lambda$ and $m_X$ where this discrepancy is $0\sigma$.  The dark gray regions contain points where the 
theoretical value of $a_\mu$ is at least $2\sigma$ larger than the 
experimental value.  The unshaded regions show points where the experimental 
value of $a_\mu$ remains at least $2\sigma$ larger than the theoretical value.}
\label{fig:lambda_allowed}
\end{center}
\end{figure*}
As we expect from Fig.~\ref{fig:a_mu}, $\lambda$ is constrained to smaller 
values in Cases I and II than in III and IV.  Also, we note that in Case III, the contribution to $a_\mu$ is proportional to $m_X$, which suppresses it for low values of $m_X$.

We also show the contribution to $a_\mu$ as functions of $m_X$ with 
$\lambda=\epsilon e=0.06$ in Case II with $Y$ identified as the muon in Fig. ~
\ref{fig:2prime}.  Also shown are allowed values of $\epsilon$ as function of 
$m_X$.
\begin{figure*}
\begin{center}
\begin{tabular}{cc}
\resizebox{80mm}{!}{\includegraphics{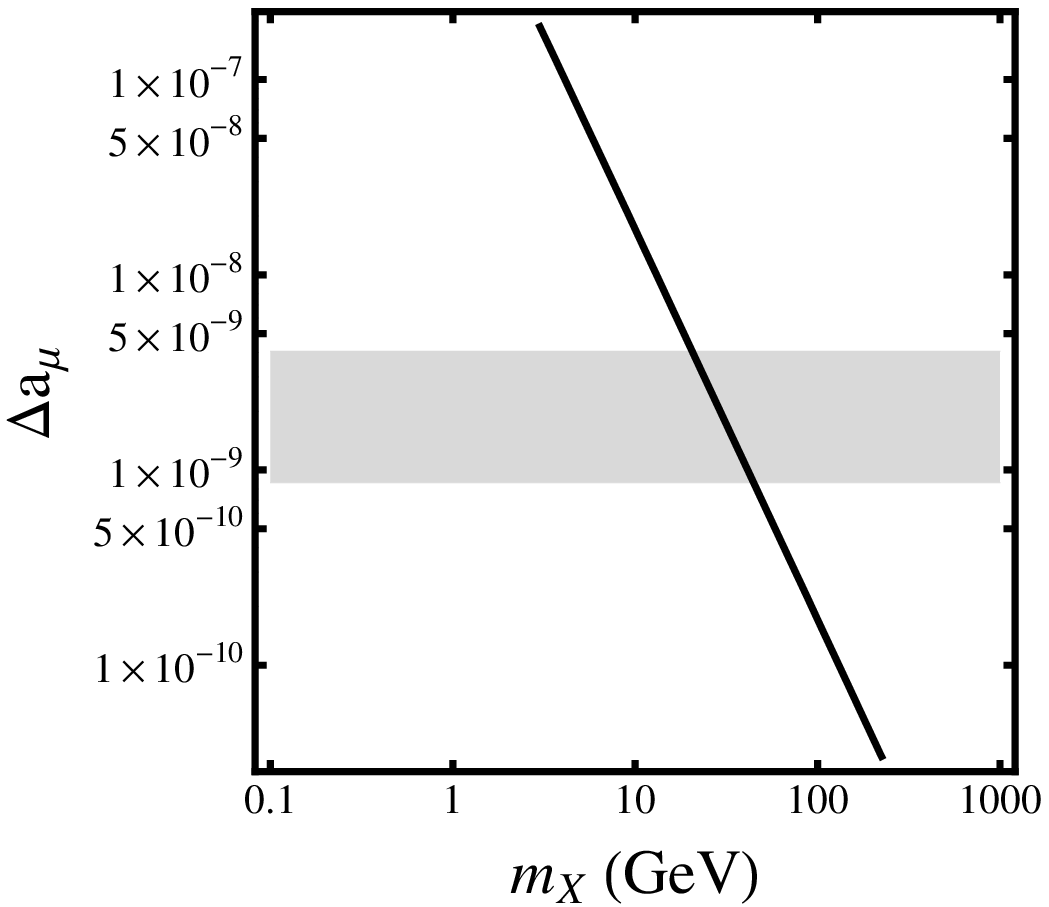}} &
\resizebox{80mm}{!}{\includegraphics{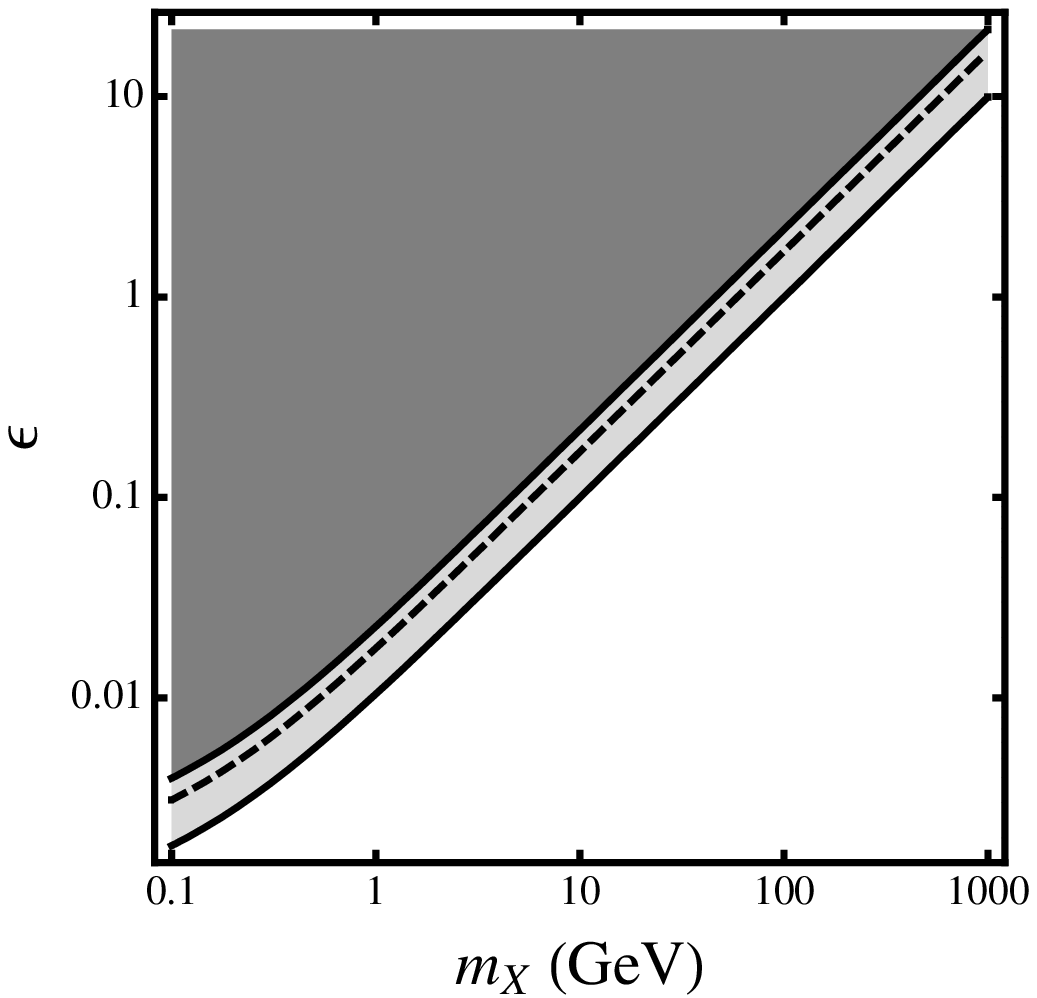}} 
\end{tabular}
\caption{Left: Contribution to $a_\mu$ as functions of $m_X$ for 
$\lambda=\epsilon e=0.06$ or $\epsilon\simeq 0.2$ in Case II with $Y$ 
identified as the muon.  The light gray band shows values of $a_\mu$ within the 
range $\Delta a_\mu\pm 2\sigma$.  Right: allowed values of $\epsilon$ as a 
function of $m_X$ in the same case.  The light gray bands indicate values of 
$\epsilon$ and $m_X$ for which the discrepancy between the theoretical and 
experimental values of $a_\mu$ is less than $2\sigma$.  The dark gray regions 
contain points where the theoretical value of $a_\mu$ is at least $2\sigma$ 
larger than the experimental value.  The unshaded regions show points where the 
experimental value of $a_\mu$ remains at least $2\sigma$ larger than the 
theoretical value.}
\label{fig:2prime}
\end{center}
\end{figure*}

We note that a fermionic $X$ (Cases III and IV) can be more strongly coupled to 
muons without violating experimental constraints on $a_\mu$ if its mass is much 
smaller than that of $Y$.  If $Y$'s are observed at the Tevatron or at the LHC, 
their decay widths can be compared with their contribution to $a_\mu$ to help determine their spin.
\end{section}

%%%%%%%%%%%%%%%%%%%%%%%%%%%%%%%%%%%%%%%%%%%%%%%%%%%%%%%%%%%%%%%%%%%%%%%%%%%%%%%

\begin{section}{Conclusions}\label{sec:conclusion}
The experimental value of $a_\mu$ and its value in the 
SM currently differ at the $3.1\sigma$ level.  This could be a sign of physics 
beyond the SM.  Hidden sectors that couple to muons can provide an explanation 
of this deviation.  In particular, situations in which the muon is coupled to 
particles that are charged under both the SM and a hidden symmetry group, $G$, 
and to particles only charged under $G$ could give rise to a nonzero 
$\Delta a_\mu$.  These particles could also be found in collider experiments 
and measurements of their spins and couplings could shed light on the 
possibility that they contribute significantly to $a_\mu$.

The spins of the hidden or mixed particles that couple to the muon greatly affect 
the structure of their contributions to $a_\mu$.  In particular, when a fermionic SM singlet is coupled to the muon with a bosonic connector, the constraints on the coupling strength from $a_\mu$ are less severe for SM singlet masses less that about $100~{\rm GeV}$.  In this way, it is easier to ``hide" a light fermionic SM singlet that couples to the muon than a bosonic one.

It is also worth considering whether couplings of the form of 
Eq.~\ref{eq:schem_lag}, in the case where $X$ is a dark matter candidate, could 
be responsible for the recent excesses seen in cosmic ray positrons seen by the 
PAMELA experiment \cite{Adriani:2008zr}.  Depending on the values of 
$\lambda_L$ and $\lambda_R$, the  dominant annihilation channel for $X$'s could 
be $XX\to \mu^+\mu^-$ through t-channel $Y$ exchange.  Dark matter decays into a 
pair of muons are seen to fit the positron data reasonably well (modulo boost 
factors) \cite{Meade:2009iu}, while the muons are kinematically constrained 
from producing baryons and so would not violate experimental limits on the 
antiproton fraction of cosmic rays.  Future work will study this in more 
detail.

A proposed muon $(g-2)$ experiment hopes to reduce the current experimental 
error on $a_\mu$ by a factor $\sim 4$ \cite{Hertzog:2008zz}.  The uncertainty 
on the difference between the theoretical and experimental values would then be 
dominated by the theoretical errors.  Such a measurement would help to 
determine the significance of the deviation between experimental and 
theoretical values of the muon's anomalous magnetic moment which is a powerful 
probe of physics beyond the SM.

\end{section}

%%%%%%%%%%%%%%%%%%%%%%%%%%%%%%%%%%%%%%%%%%%%%%%%%%%%%%%%%%%%%%%%%%%%%%%%%%%%%%%

\begin{acknowledgments}
The author would like to thank J.~L.~Rosner, Q.-H.~Cao, R.~J.~Hill, D.~Krop, A.~M.~Thalapillil, and P.~Draper for 
discussions and helpful suggestions.  This work was supported in part by the 
United States Department of Energy under Grant No. DE-FG02-90ER40560.
\end{acknowledgments}

%%%%%%%%%%%%%%%%%%%%%%%%%%%%%%%%%%%%%%%%%%%%%%%%%%%%%%%%%%%%%%%%%%%%%%%%%%%%%%%

\appendix

\section{$Y$ Pair Production in $e^+e^-$ Collisions}
\subsection{Fermionic $Y$ Production Cross Section}\label{sec:fermprod}
\begin{figure*}
\begin{center}
\resizebox{60mm}{!}{\includegraphics{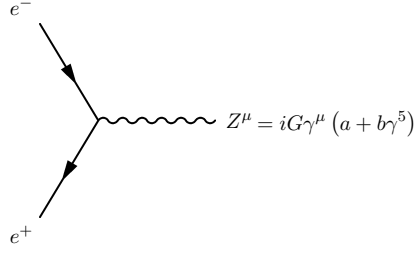}}
\caption{Feynman rules for the $e^+-e^--Z$ vertex (and the $\mu^+-\mu^--Z$ vertex).  $G=e/\left(\sin\theta_W\cos\theta_W\right)$, $a=1/4-\sin^2\theta_W$, and $b=1/4$ with the weak mixing angle $\sin^2\theta_W=0.231$.}
\label{fig:feyneez}
\end{center}
\end{figure*}

For a fermionic $Y$, we consider the case where its representation under $SU(2)_L\times U(1)_Y$ is the same as that of the muon.  That is, $Y_L$ is a doublet under $SU(2)_L$ with hypercharge $-1$ while $Y_R$ is an $SU(2)_L$ singlet with hypercharge $-2$.  Its couplings to the photon and $Z$ boson are then the same as the muon's.   The Feynman rule for the electron's (which is the same as the muon's) coupling to the $Z$ boson is shown in Fig.~\ref{fig:feyneez}.
The cross section, ignoring the width of the $Z$, for $e^+e^-\to Y\bar{Y}$ is easily found to be
\begin{align}
\sigma\left(e^+e^-\to Y\bar{Y}\right)&=\frac{2\pi\alpha^2}{s} \beta\left\{\left[1+\frac{2 G^2 a^2}{e^2}\left(\frac{s}{s-m_Z^2}\right)\right]\left(1-\frac{1}{3}\beta^2\right)\right.
\\
&\quad\quad\quad\quad\left.+\frac{G^4 \left(a^2+b^2\right)}{e^4}\left(\frac{s}{s-m_Z^2}\right)^2\left(a^2-\frac{1}{2}\left(a^2-b^2-\frac{1}{3}\right)\beta^2\right)\right\}
\end{align}
where $G$, $a$, and $b$, as seen in Fig.~\ref{fig:feyneez}, are expressed in terms of the electron's charge and the weak mixing angle $\theta_W$ as
\begin{align}
G&=\frac{e}{\sin\theta_W\cos\theta_W}~~~,
\label{eq:G}
\\
a&=\frac{1}{4}-\sin^2\theta_W~~~,
\\
b&=\frac{1}{4}~~~.
\label{eq:b}
\end{align}

\subsection{Scalar $Y$ Production Cross Section}\label{sec:scalarprod}
\begin{figure*}
\begin{center}
\begin{tabular}{cc}
\resizebox{60mm}{!}{\includegraphics{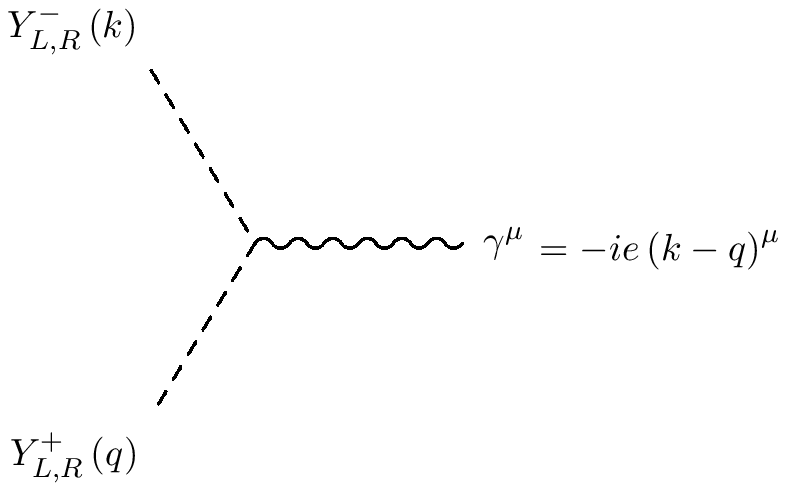}} &
\resizebox{60mm}{!}{\includegraphics{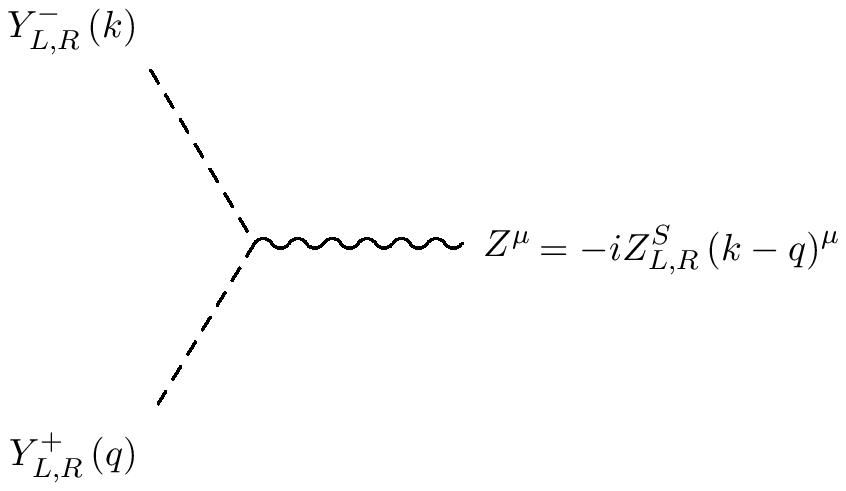}} \\
\end{tabular}
\caption{Feynman rules for a scalar $Y$.  All momenta are running into the graphs.  We use $Z^S_L=e\left(1-2\sin^2\theta_W\right)/\left(2\sin\theta_W\cos\theta_W\right)$ and $Z^S_R=-e\tan\theta_W$.}
\label{fig:feynscalar}
\end{center}
\end{figure*}

For a scalar $Y_L$ we assume that it couples to the $Z$ boson in a gauge invariant way with a strength equal to that of the left-handed muon.  We assume analogously for a scalar $Y_R$.  The Feynman rules for these $Y_L-Y_R-\gamma$ and $Y_L-Y_R-Z$ couplings are shown in Fig.~\ref{fig:feynscalar}.
We then obtain
\begin{align}
\sigma\left(e^+e^-\to Y_{L,R}^+Y_{L,R}^-\right)=\frac{\pi\alpha^2}{s} \left(\frac{\beta^3}{3}\right)\left\{1+\frac{2 G a}{e^2}Z_{L,R}^S \left(\frac{s}{s-m_Z^2}\right)+\frac{G^2 \left(a^2+b^2\right)}{e^4}\left(Z_{L,R}^S\right)^2\left(\frac{s}{s-m_Z^2}\right)^2\right\}
\end{align}
where we have again ignored the width of the $Z$.  $G$, $a$, and $b$ are as in Eqs.~\ref{eq:G}-\ref{eq:b} and
\begin{align}
Z^S_L&=\frac{e\left(1-2\sin^2\theta_W\right)}{2\sin\theta_W\cos\theta_W}~~~,
\\
Z^S_R&=-e\tan\theta_W~~~.
\end{align}

\subsection{Vector $Y$ Production Cross Section}\label{sec:vectorprod}
\begin{figure*}
\begin{center}
\begin{tabular}{ccc}
\resizebox{60mm}{!}{\includegraphics{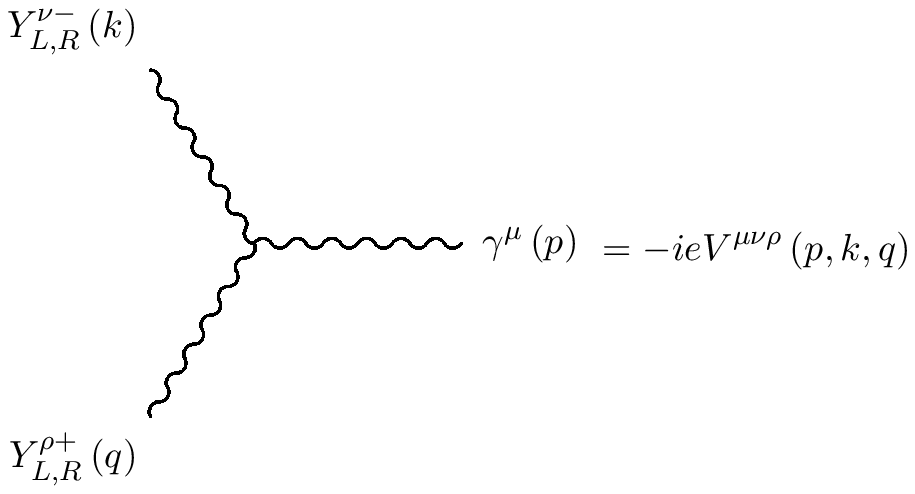}} &
\resizebox{60mm}{!}{\includegraphics{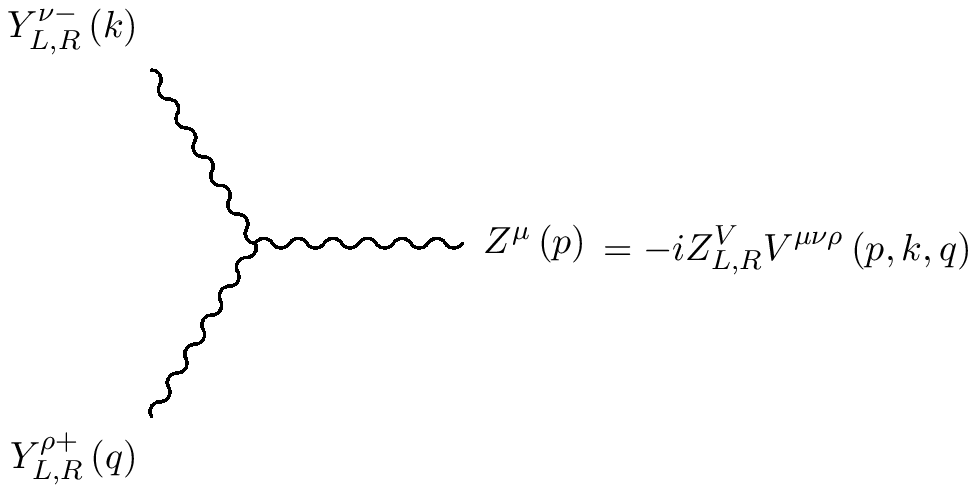}} \\
\end{tabular}
\resizebox{60mm}{!}{\includegraphics{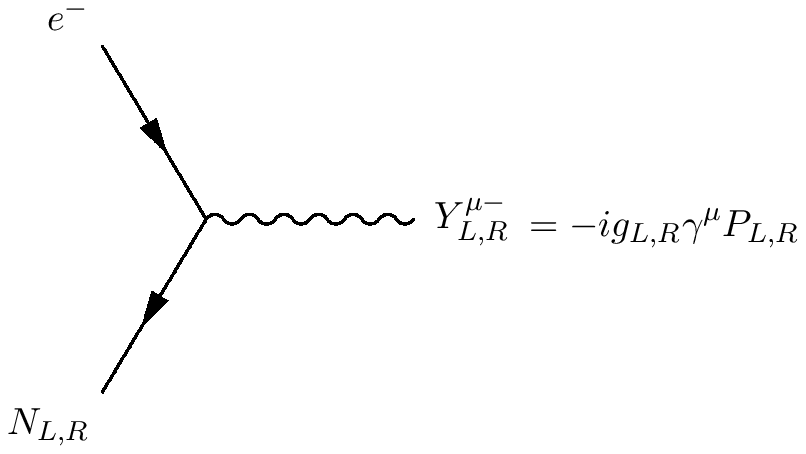}}
\caption{Feynman rules for a vector $Y$.  $V^{\mu\nu\rho}\left(p,k,q\right)=\left[\left(p-k\right)^\rho g^{\mu\nu}+\left(k-q\right)^\mu g^{\nu\rho}+\left(q-p\right)^\nu g^{\mu\rho}\right]$ and all momenta are running into the graphs.  The photon interaction is fixed by demanding electromagnetic gauge invariance.  The values of $g_L$ and $Z_{L,R}^V$ are determined by requiring that they that keep $Y$ pair production unitary.  They are given in Eq.~\ref{eq:gZsoln}.}
\label{fig:feynvector}
\end{center}
\end{figure*}

To properly determine the cross section for vector $Y$ pair production we need to introduce new states to insure unitarity is not violated.  As mentioned in Sec.~\ref{sec:collconstr}, we assume that there is an electrically neutral fermion that couples to $Y_{L,R}^\nu$ and $e_{L,R}$ which we call $N_{L,R}$.  We write the Feynman rules for the interactions of $Y_{L,R}^\nu$ with $e^\pm$, $Z$, and $N_{L,R}$ in Fig.~\ref{fig:feynvector}.  The coupling strengths $g_L$ and $Z_{L,R}^V$ will be chosen so that the cross section for $e^+e^-\to Y_L^{\nu+} Y_L^{\nu-}$ remains finite as $\sqrt{s}\to\infty$.

We write the cross section for the production of pair of vector $Y_L$s as
\begin{align}
\frac{d\sigma}{d\cos\theta}\left(e^+e^-\to Y_L^{\nu+} Y_L^{\nu-}\right)=\frac{\pi\alpha^2}{16}\frac{\beta}{s}\sum \left|{\cal M}_{ij}\right|^2
\end{align}
where $\theta$ is the center-of-mass scattering angle.  The squared matrix elements are
\begin{align}
\left|{\cal M}_{NN}\right|^2&=\frac{4g_L^4}{e^4}\left(\frac{t}{t-m_{N_L}^2}\right)^2 F_t\left(t, s\right)~~~,
\\
\left|{\cal M}_{\gamma\gamma}\right|^2&=F_s\left(t, s\right)~~~,
\\
\left|{\cal M}_{ZZ}\right|^2&=\frac{G^2\left(a^2+b^2\right)\left|Z_L^V\right|^2}{e^4}\left(\frac{s}{s-m_Z^2}\right)^2 F_s\left(t, s\right)~~~,
\\
\left|{\cal M}_{Z\gamma}\right|^2&=\frac{2 G a~{\rm Re}(Z_L^V)}{e^2}\left(\frac{s}{s-m_Z^2}\right) F_s\left(t, s\right)~~~,
\\
\left|{\cal M}_{NZ}\right|^2&=-\frac{2G\left(a+b\right){\rm Re}(Z_L^V) g_L^2}{e^4}\left(\frac{s}{s-m_Z^2}\right)\left(\frac{t}{t-m_{N_L}^2}\right) F_{st}\left(t, s\right)~~~,
\\
\left|{\cal M}_{N\gamma}\right|^2&=-\frac{2g_L^2}{e^2}\left(\frac{t}{t-m_{N_L}^2}\right) F_{st}\left(t, s\right)~~~,
\end{align}
where $s$ and $t$ are the usual Mandelstam variables and we have defined the functions
\begin{align}
F_t\left(t, s\right)&=2\left(\frac{s}{m_{Y_L}^2}\right)+\frac{1}{2}\beta^2\sin^2\left[\theta\left(t\right)\right]\left[\left(\frac{s}{t}\right)^2+\frac{1}{4}\left(\frac{s}{m_{Y_L}^2}\right)^2\right]~~~,
\\
F_s\left(t, s\right)&=\beta^2\left\{16\left(\frac{s}{m_{Y_L}^2}\right)+\sin^2\left[\theta\left(t\right)\right]\left[\left(\frac{s}{m_{Y_L}^2}\right)^2-4\left(\frac{s}{m_{Y_L}^2}\right)+12\right]\right\}~~~,
\\
F_{st}\left(t, s\right)&=16\left(1+\frac{m_{Y_L}^2}{t}\right)+8\beta^2\left(\frac{s}{m_{Y_L}^2}\right)+\frac{1}{2}\beta^2\sin^2\left[\theta\left(t\right)\right]\left[\left(\frac{s}{m_{Y_L}^2}\right)^2-2\left(\frac{s}{m_{Y_L}^2}\right)-4\left(\frac{s}{t}\right)\right]~~~.
\\
\end{align}
We relate $\theta$ and $t$ through
\begin{align}
\sin^2\left[\theta\left(t\right)\right]=-\frac{4}{\beta^2}\left[\left(\frac{t-m_{Y_L}^2}{s}\right)^2+\frac{t}{s}\right]~~~.
\end{align}

Unitarity will determine the values of $g_L$ and $Z^V_L$.  Requiring that the coefficients of $(s/m_{Y_L}^2)^2\sin^2\theta$ and of $(s/m_{Y_L}^2)$ vanish as $\sqrt{s}\to\infty$ gives
\begin{align}
g_L^4+2\left[e^4+G^2\left(a^2+b^2\right)\left|Z_L^V\right|^2+2 e^2G a~{\rm Re}(Z_L^V)\right]-2g_L^2\left[G\left(a+b\right){\rm Re}(Z_L^V)+e^2\right]=0~~~,
\end{align}
while setting the coefficient of $(s/m_{Y_L}^2)\sin^2\theta$ to zero as $\sqrt{s}\to\infty$ implies
\begin{align}
2\left[e^4+G^2\left(a^2+b^2\right)\left|Z_L^V\right|^2+2 e^2G a~{\rm Re}(Z_L^V)\right]-g_L^2\left[G\left(a+b\right){\rm Re}(Z_L^V)+e^2\right]=0~~~.
\end{align}
These two equations are satisfied by
\begin{align}
g_L^2=\frac{2e^2b}{b-a}~~~,~~~Z_L^V=\frac{e^2}{G\left(b-a\right)}~~~.
\label{eq:gZsoln}
\end{align}
We note that the contribution due to $Z$ boson exchange cannot by itself cancel that from photon exchange unless the $Z$ coupling to electrons is vector-like (which it is not).  These conditions allow us to determine $\sigma\left(e^+e^-\to Y_L^{\nu+} Y_L^{\nu-}\right)$ as a function of $m_{Y_L}$ and $m_{N_L}$.

$\sigma\left(e^+e^-\to Y_R^{\nu+} Y_R^{\nu-}\right)$ is obtained from $\sigma\left(e^+e^-\to Y_L^{\nu+} Y_L^{\nu-}\right)$ by taking $b\to -b$.

%%%%%%%%%%%%%%%%%%%%%%%%%%%%%%%%%%%%%%%%%%%%%%%%%%%%%%%%%%%%%%%%%%%%%%%%%%%%%%%

\end{document}